# Comparison of Two-Moment and Three-Moment Bulk Microphysics Schemes in Thunderstorm Simulations over Indian Subcontinent

Chandrima Mallick[1,2], Ushnanshu Dutta[3], Moumita Bhowmik[1], Greeshma M. Mohan[4], Anupam Hazra[1,3*], S. D. Pawar[1], and Jen-Ping Chen[3]

[1]Indian Institute of Tropical Meteorology, Ministry of Earth Sciences, Pune 411008, India.

[2]Savitribai Phule Pune University, Pune, India.

[3]Department of Atmospheric Sciences, National Taiwan University, Taipei, Taiwan 106.

[4]National Center for Medium Range Weather Forecasting, Ministry of Earth Sciences, Noida, India.

**Abstract**

We have performed three-dimensional thunderstorm real simulations using the two-moment and three-moment bulk microphysics schemes in the Weather Research and Forecasting (WRF) model. We have analyzed three cases to understand the potential differences between the double-moment (Morrison-2M) and National Taiwan University triple-moment (NTU-3M) microphysics parameterizations in capturing the characteristics of lightning events over the Indian subcontinent. Despite general resemblances in these schemes, the simulations reveal that there is a distinct difference in storm structure, cloud hydrometeors formation, and precipitation. The lightning flash counts from the in situ lightning detection network (LDN) are also used to compare the simulation of storms. The Lightning Potential Index (LPI) is computed for Morrison-2M and NTU-3M microphysics schemes and compared it with the Lightning Detection Network (LDN) observation. In most cases, the Morrison-2M shows more LPI than the NTU-3M scheme. Both the schemes also differ in simulating rainfall and other thermodynamical, dynamical, and microphysical parameters in the model. Here, we have attempted to identify the basic differences between these two schemes, which may be responsible for the discrepancies in the simulations. . In particular, the Morrison-2M produced much higher surface precipitation rates. The effects on the size distributions cloud hydrometeors between two microphysical schemes are important to simulate the biases in the precipitation and lightning flash counts. The inclusions of ice crystal shapes are responsible for many of the key differences between the two microphysics simulations. Different approaches in treating cloud ice, snow, and graupel may have an impact on the simulation of lightning and precipitation. In fine, the results show that the simulation of characteristics of lightning events is sensitive to the choice of microphysical parameterization schemes in numerical weather prediction models.

*Keywords:* Thunderstorms/Lightning, Cloud microphysics, WRF

**Highlights**

• We simulate three thunderstorm cases using 3M and 2M bulk microphysics schemes.

• Simulations using 3M microphysics scheme reproduce better storm characteristics as compared to 2M scheme.

• Surface rainfall and lightning biases are less in 3M microphysics scheme.

# 1. Introduction:

Lightning is a highly localized, short-time scale, lethal, and disastrous weather event which occurs associated with thunderstorms. It is one of the major natural calamities which cause casualties, forest fire, severe damages to infrastructures, electrical structures, living and non-living objects etc. The lightning leads to the production of $NO_x$ which may subsequently cause the production of ground level ozone, a major air pollutant. Mohan et al. (2021) from observational data (i.e., Tropical Rainfall Measuring Mission-Lightning Imaging Sensor, Cecil et al., 2014) found that the lightning strikes more over the tropical land than that of the ocean during pre-monsoon (March-April-May). Over India, higher density of lightning flashes is observed over the eastern, southern peninsula and northern parts. Tyagi (2007) also found that these regions (e.g., Assam, West Bengal, Jammu) have highest annual frequency of lightning (i.e., 100-120 days). Mahapatra et al., (2018), analyzed the casualties caused by severe weather conditions, e.g., extreme precipitation, lightning, tropical cyclones cold wave, heat wave, etc. during 2001 to 2004. They found that most of the death percentage (39.8% of deaths per million) are caused by lightning. Also, they found that there is an increasing trend in death caused by lightning during the study period.

Lightning is known to have strong microphysical in origin, which is responsible for charge separation processes and generation of electric field (Adamo et al., 2007). Previous studies (Reynolds et al., 1957; Takahashi, 1978; Jayaratne et al., 1983; Mansell et al., 2005; Saunders, 2008; Saunders et al., 1991) have shown that the charge separation can be caused by the rebounding collisions between graupel particles and cloud ice crystals, in the presence of supercooled liquid water. The charge separation leads to the typical charge distribution with negative charges between the temperature regions −25°C and −10°C (graupel) and positive charges at the upper part of the storm clouds (cloud ice). Studies by several researchers have proposed that more aerosols, which act as cloud condensation nuclei (CCN) modulate the dynamics of the cloud development and thereby lightning (Tao et al., 2012; Kawecki et al., 2018; Dayeh et al., 2021; Bathlott et al., 2022). In the convective environment, the CCN basically produce smaller cloud drop size, reducing collision-coalescence efficiency for warm rain and helps smaller cloud droplets to uplift below freezing level to form cloud ice (Posselt and Lohmann 2008; Freud et al., 2008, Bathlott et al., 2022). The observations pointed out that dust devils can have intense electrification (Franzese et al., 2018), which is revealed from the measurement (Dani et al., 2003) at Roorkee, India. The potential gradient due to dust devils was recorded and the increasing wind speeds modulates the systematic deviation of potential conductivity and its electrical agitation (Dani et al., 2003).

    Bulk microphysics schemes play an important role in operational numerical weather prediction (NWP) and atmospheric research. Simulations of severe weather events including lightning/thunderstorms have largely benefited due to the increasing computational resources in recent years, where bulk microphysical scheme parameterize cloud and precipitation. However, discussion arises regarding whether to allocate resources primarily towards refining the intricacies of physics within deterministic simulations or towards improving spatial and temporal resolution. This underscores the importance of comprehensively grasping the benefits that enhanced physics can bring to severe weather or deep convection simulations. Currently, weather and climate simulations are increasingly moving towards convection-resolving scales for dynamical downscaling of large-scale simulations. In these simulations, the cumulus parameterization scheme can be turned off, allowing all clouds and precipitation to be generated solely by the microphysical parameterization (Weverberg et al., 2014). A bulk particle size distribution for each hydrometeors type is considered in a typical bulk microphysics scheme. A recent advancement of multi-moment microphysics scheme has been considered to relate more prognostic model variables by generating the number concentration of the cloud

and precipitation particles (Morrison et al., 2005, 2009; Chen and Tai 2016, Tsai and Chen 2020). Although employing multi-moment schemes demands significantly greater computational resources, several studies indicate a distinct advantage in utilizing these schemes for simulating deep convective precipitation (Morrison et al., 2009, Tao et al., 2014).

Double-moment bulk microphysics schemes are currently used operationally (Milbrandt and Morrison, 2016; Benjamin et al., 2016; Vié et al., 2016), while triple-moment schemes are emerging as prominent alternatives (Chen & Tsai, 2016; Luo et al., 2018; Tsai & Chen, 2020; Mansell et al., 2020; Milbrandt et al., 2021), offering increased flexibility in representing hydrometeor size distributions and computing microphysical process rates. Despite these advancements, nearly all current bulk microphysics schemes still lack explicit representation of mixed-phase hydrometeors (Cholette et al., 2020), which remains a significant limitation in parameterizing various crucial microphysical processes as mentioned by Frick et al., (2013). The characterization of ice crystal shape and density (Chen and Tsai, 2016) and the refreezing process are frequently disregarded (Frick et al., 2013). However, while the studies mainly concentrate on work within an idealized framework, real-case simulations using NWP models are scarce (Morrison and Milbrandt, 2011; Baldauf et al., 2011). Therefore, a comprehensive investigation into the impact on cloud and precipitation dynamics, alongside an in-depth process-based evaluation against observational evidence, is essential to ascertain the added value of employing a more complex three-moment microphysics scheme over a two-moment counterpart. In this study we perform sensitivity tests to evaluate the impact of the three-moment microphysical scheme on the simulation of convective events (lightning) over Indian sub-continent. In this study, we simulated all these occurrences using a detailed climate model (WRF) with both two-moment and three-moment microphysics, then compared the results with available data from lightning detection networks, radar reflectivity, and rain gauges.

The details of the regional climate model (WRF) and its configurations in the present simulations are presented in the next section. The observational datasets (in situ LDN data over India, Rainfall), re-analysis product used here are presented in the section 3. Section 4 provides the results of the sensitivity experiments and we summarize the study in the Section 5.

**2. Data and Methodology:**

**2.1. Datasets used:**

The fifth generation of the European Centre for Medium-Range Weather Forecasts (ECMWF) reanalysis product, ERA5 (Hersbach et al., 2020), is considered to elucidate the synoptic features on the selected days for this study. It is based on ECMWF's Earth System model, the Integrated Forecasting System (IFS), and is a better choice among reanalysis products (Pokhrel et al., 2020, Dutta et al., 2022). The relative humidity, mean sea level pressure and wind (zonal and meridional components) data analyzed from ERA5. We have used rainfall data from TRMM Multi-Satellite Precipitation Analysis TMPA (3B42v7, (Huffman et al., 2007, 2010) to validate the model simulated rainfall. Lightning Detection Network (LDN) observations are used for assessing the nature of the model simulated lightning over the study regions. This network comprises of 20 Earth Network Lightning Sensors (ENLS) those are capable of detecting both intra-cloud (IC) and cloud-to-ground (CG) lightning flashes. The details of LDN can be found in Mohan et al., (2021).

**2.2. Model description:**

This study utilizes the Advanced Research WRF model (ARW), version 3.8.1 (Skamarock et al., 2008). WRF is a mesoscale atmospheric model that is fully compressible and non-hydrostatic, employing a Eulerian mass dynamical core. Its vertical coordinate system adopts a terrain-following hydrostatic pressure coordinate, while it employs the Arakawa-C grid horizontally.

The model is configured with two nested domains (Fig. S1), having horizontal resolutions of 3, and 1 km, and there are 45 vertical levels up to 50 hPa. There are three lightning cases selected for the present study, viz., Case1: 03rd May 2019, Case2: 25th May 2019 and Case3: 28th May 2019. The mother domain is kept same for the simulations, while the innermost domain (1 km) is different for Case3. For the first two cases, the innermost domain covers the Eastern coast of India (Fig. S1a) and for the third case it covers the southern peninsula of India (Fig. S1b). Initialization of the model involves National Centers for Environmental Prediction (NCEP) FNL (Final) 6-hourly data with a resolution of 0.25°x0.25°. The model is integrated up to 30 hours with output stored hourly, considering the first 6 hours as the model spin-up period.

The WRF model in this study incorporates various physics parameterization schemes. Yonsei University Scheme for the planetary boundary layer, a new version of the Rapid Radiative Transfer Model (RRTMG) for short and longwave radiation are considered. We have used revised MM5 Monin-Obukhov scheme for surface layer, and for land surface Unified Noah land-surface model is employed. Explicit convection is employed for the both cloud-resolving domains. Ice microphysics plays an important role on the simulation of weather events associated with deep convection, such as lightning/thunderstorms. However, highly simplified representations of spherical ice crystal shape with fixed density can have adverse effects on charge separation in thunder clouds. Therefore, a multi-moment bulk microphysical scheme with four-ice (pristine ice, aggregate, graupel, and hail) categories by Chen and Tsai (2016) (hereafter NTU scheme) has been implemented into the WRF. The NTU scheme in WRF introduces several improvements over existing bulk microphysics schemes. It employs a triple-moment closure method for all ice categories, allows for the evolution of ice crystal shape (Chen and Tsai, 2016), predicts apparent densities of various hydrometeors. It also fully couples crystal shape, apparent density, and fall speed for solid hydrometeors following the theoretical parameterization by Mitchell and Heymsfield (2005).

**2.3. Double and Triple Moment Microphysics scheme:**

The three key elements (Chen and Tsai 2004) of a bulk microphysical parameterization are i) Size distribution function n(D), ii) moments ($M_k$) which controls the conversion to/from n(D) and iii) Growth Kernel (K), which decides the time rate of change of n(D).

$$M_k = \int D^k n(D) dD \tag{1}$$

$$\frac{dM_k}{dt} = \int K_k n(D) d(D), where, K_k = \frac{dD^k}{dt} \tag{2}$$

If we assume the gamma size distribution i.e., $n(D) = N_0 \, D^\alpha . \exp(-\lambda D)$

then,

$$M_k = \int D^k n(D)dD = \frac{N_0 \Gamma(k+\alpha+1)}{\lambda^{k+\alpha+1}} \; ; \; \Gamma(a+1) = a\Gamma(a) \tag{3}$$

Therefore, the moments considered in microphysical parametrization schemes are as follows

$$\text{Zeroth Moment:} \quad M_0 = N = \int n(D)dD \; ; \text{N: number} \tag{4}$$

$$\text{First Moment:} \; M_1 = N.\overline{D} = \int D.n(D)dD \; ; \; \overline{D} = \text{mean diameter} \tag{5}$$

$$\text{Second Moment:} \; M_2 = \frac{A}{4\pi} = \int D^2.n(D)dD \; ; \text{A: area} \tag{6}$$

$$\text{Third Moment:} \; M_3 = V\frac{6}{4\pi} = \int D^3.n(D)dD \; ; \text{V: Volume} \tag{7}$$

$$\text{Sixth Moment:} \; M_6 = Z = \int D^6 n(D)dD : \text{Z: Radar reflectivity factor} \tag{8}$$

The degree (i.e., single, double, triple, etc.) of a microphysics scheme is determined by number of moments tracked by a cloud model. For example, Morrison Microphysics scheme (Morrison et al. 2005) considers Zeroth and Third moment. Hence, it is known as 'Double' moment microphysics scheme (hereafter Morrison-2M). On the other hand, NTU scheme (Chen and Tai, 2016) considers the second moment along with zeroth and third moment and thus it is 'Triple' moment microphysics scheme (hereafter NTU-3M).

Consequently, sensitivity experiments aim to explore the role of microphysics parameterizations in simulating lightning and rainfall. Acknowledging the solid microphysical origin of lightning, the study emphasizes the relationship between lightning, dynamics, and microphysics. The study poses the question: Does the moment of the microphysics indeed have an impact on lightning and rainfall? This influence is demonstrated through the real simulations of the case study using two different microphysics parameterization schemes.

**2.4. Calculation of Lightning Potential Index (LPI):**

The LPI (J/kg) quantifies the potential for charge generation and separation within clouds, determined by factors such as vertical velocity and cloud hydrometeors (Yair et al., 2010). The LPI is calculated as follows

$$LPI = \frac{1}{v} \iiint \epsilon \omega^2 dx dy dz \tag{9}$$

Where, v= volume of air in the layer between 0 °C and − 20 °C, ω= vertical velocity (m/s), ε= Dimensionless number and value between 0 to 1 which is defined as follows

$$\epsilon = \frac{2(Q_i Q_l)^{0.5}}{(Q_i + Q_l)} \tag{10}$$

Where, $Q_i$= Ice fractional mixing ratio (kg kg$^{-1}$), $Q_l$= Mass mixing ratio of total liquid water (kg kg$^{-1}$). Further $Q_i$ is defined as

$$Q_i = q_s \left[ \frac{(q_s q_g)^{0.5}}{(q_s+q_g)} + \frac{(q_i q_g)^{0.5}}{(q_i+q_g)} \right] \tag{11}$$

Where, $q_s$, $q_i$ and $q_g$ are the mass mixing ratio (kg kg$^{-1}$) for snow, ice and graupel respectively.

**2.5. Synoptic conditions during the selected events:**

**2.5.1. Case 1: 03 MAY 2019:**

The major synoptic weather event identified at 0000 UTC of 03rd May 2019 is the cyclonic circulation over the head Bay (Fig. 1a,d). IMD reported it as an "Extremely severe cyclonic storm (ESCS)" and was named as "Fani". The peak intensity of this system is reported during the evening hours of 2nd May to early hours of 3rd May 2019 and slightly weakened thereafter by 0000 UTC. It crossed the land on the same day between 0230 to 0430 UTC. In this case, i.e., 03rd May 2019, the thunderstorms within the rain bands associated with this ESCS produced lightning. The increased amount of moisture (relative humidity greater than 80%) enhances the sustenance of the thunderstorms.. The reported CAPE value (University of Wyoming) over Kolkata is 1204 J/kg. The observed TRMM rainfall and lightning activity (discussed later) occurred mainly over the forward sector of the cyclone. Both the schemes (Morrison-2M and NTU-3M) are able to capture the synoptic conditions pretty well (Fig. 1g,j) as compared to ERA5 (Fig. 1d).

### 2.5.2. Case 2: 25 MAY 2019:

This case is featured with strong prevailing south westerlies which bring moisture to the Gangetic West Bengal (Fig. 1b,e). These strong south westerlies also converge with the north westerlies over this region. The availability of high moisture is also noted over this region. The reported CAPE value (University of Wyoming) over Kolkata is 1474 J/kg. The observed TRMM rainfall and lightning activity (discussed later) are also consistent with this zone of convection. For this event, the synoptic conditions are well simulated by the both cloud microphysical schemes (Morrison-2M and NTU-3M) (Fig. 1h,k) as compared to ERA5 (1e).

### 2.5.3. Case 3: 28 MAY 2019:

The north-westerlies along the west coast assure the availability of moisture from the Arabian Sea which favors the feeds the convection and thereby the lightning over that region (Fig. 1c,f). The reported CAPE value nearby Karaikal is 3647 J/kg. The simulated synoptic conditions by both of the cloud microphysical schemes (Morrison-2M and NTU-3M) are similar (Fig. 1i,l), but underestimated relative humidity over western Ghats as compared to ERA5 (1f).

### 3. Results and Discussions:

### 3.1. Lightning and Reflectivity:

In this study, we focused three specific lightning events for a sensitivity analysis to understand the advantages of the complex microphysics over the simple one in simulating the convective weather events. The storm indices, obtained from the nearest radio-sounding observation (University of Wyoming) for each of the three cases are listed in Table 1. The possibility of the strong convection in the Case-1 and Case-2 is well marked by all these storm parameters, whereas in Case-3 Total totals index and CAPE supports the possibility of the thunderstorm activity. This is further confirmed with the LDN data which indicate the areas of strong convection and resulting in lightning flashes. Figure 2 displays the spatial distribution of lightning flashes observed from in-situ LDN measurements for all the three cases. The widespread lightning activity is noticed over the East coast of India for the first two cases (Fig. 2a, b) while the southern peninsula is stricken by heavy lightning for the third case (Fig. 2c The lightning potential index (LPI) over lightning prone zone from the simulations for the three cases are listed in Table 2 and it is found that LPI were more in Case-2 and Case-3 than Case-1 for NTU-3M, which is seen in the observed lightning flash counts from LDN (Figure 2). On the other hand, Morrison-2M shows more for Case-1 and least for Case-2 (Table 2). The Case-1 shows a difference in the area-averaged LPI from the Morrison-2M and NTU-3M simulations. This is further investigated through the analysis of the time evolution of the simulated maximum reflectivity.

The Fig. 3 demonstrates the evolution of maximum reflectivity (dBZ) from both the simulations (2M and 3M) correspond to 01, 02, 06, 09 ,11, 12 UTC (Fig. 3a-f for Morrison 2M, and Fig. 3g-l for NTU-3M) for the Case-1 1.e., 03rd May 2019. The time steps are chosen according to the availability of the observed radar reflectivity from IMD (Figure S2). Both the models could capture the spatial pattern of the reflectivity, especially the reflectivity over the eyewall region and the curved band of clouds of the cyclone. Though there is an over-estimation in simulated reflectivity, the pattern and evolution captured well. Significant variation in propagation of maximum reflectivity is not noticed between Morrison-2M and NTU-3M. Similar analysis for Case-2 and Case-3 are not reported because of unavailability of observation.

**3.2. Dynamics and cloud hydrometeors:**

The temporal evolution of vertical profile of vertical velocity (averaged over respective domains mentioned in Figure 2) from both the simulations for all the cases are shown in Figure 4. It is noticed that experiments with Morisson-2M simulates higher vertical velocity than that of NTU-3M for Case1 and Case3. On the other hand, the vertical velocity is more in NTU-3M for Case2 (Figure 4e). The difference is more in the middle to higher tropospheric levels. For Case-1 the substantial vertical velocity is noticed in both the experiments throughout the time steps (Fig. 4a, d), since it is associated with a synoptic scale extreme weather event. The maxima of vertical velocity are noticed at same time for Morisson-2M (Fig. 4a) and NTU-3M (Fig. 4d). For Case-2 both the experiments (Fig. 4b, e) demonstrated overall less vertical velocity as compared to Case-1. The peak of the vertical velocity is around 12 UTC from both the simulations. The peak value is also higher in NTU-3M (Fig. 4e) than Morisson-2M (Fig. 4b). For case-3 the peak is noticed at 15 UTC from both the simulations. In this case also the magnitude of vertical velocity is more in Morisson-2M (Fig.4c) than NTU-3M (Fig. 4f). Maximum vertical velocity (w-max) is an important parameter for lightning, which has a major role in controlling the interaction between hydrometeors. To understand how this complex microphysics parameterization alters the dynamics of the system, the diurnal variation of w-max averaged over lightning prone regions (see Fig.2) for each case (Fig.5a-c) is examined. The peak of w-max is noticed around 12 UTC for Morisson-2M, whereas around 07UTC for NTU-3M for Case-1. On the other hand, for Case-2 the peak of w-max appears at 12 UTC (10 UTC) for Morrison-2M (NTU-3M) scheme (Fig. 5b). For Case-3 also the peak of w-max lags behind that of Morisson-2M (Fig.5c). The peak of w-max from NTU-3M is in well agreement with the peak of diurnal variation of lightning flash counts (Fig. 5d) for Case-1 and Case-2, with the peak at 6 UTC and 10 UTC respectively.

The time-height plot of domain averaged cloud hydrometeors (i.e., sum of cloud ice, cloud liquid water, snow and graupel) and rainwater mixing ratio for all the cases is shown in Figure 6. Results show that distribution of cloud hydrometeors is sensitive to the choice of microphysics parameterization scheme. The vertical extent of mass of the cloud hydrometeors or is more (i.e., stronger development) in Morisson-2M (Fig. 6a, c, e) than NTU-3M (Fig. 6b, d, f) for all the cases. For Case-1, the rain water mixing ratio is more in the lower tropospheric levels for NTU-3M (Fig. 6b) than Morisson-2M which is almost similar for other two cases. The maxima of cloud hydrometeors are also consistent with vertical velocity (Fig. 5) which signifies the linkage between the dynamics and production of cloud hydrometeors.

To explore the underlying factors which contribute to the variation of cloud hydrometeor profile in the two simulations, the vertical and time averaged cloud hydrometeors (e.g., cloud ice, snow, graupel, cloud water, and rainwater) are analyzed for all the cases from both the experiments (Figs. 7-9). Results show that the NTU-3M simulates more cloud ice (Fig. 7b, d, f) for each case than Morisson-2M (Fig. 7a, c, e). Contrastingly, the snow mixing ratio (Fig. 7g-l) is much higher in Morisson-2M (Fig.

7g, i, k) than that of NTU-3M (Fig. 7h, j, l) for all the cases. The graupel mixing ratio is also (Fig. 8) higher in Morisson-2M than NTU-3M for all the cases. There are remarkable differences in the distribution of cloud liquid water mixing ratio between Morrison-2M and NTU-3M. The difference is more pronounced for Case-1 than other two cases. NTU-3M shows more cloud liquid water than Morrison-2M (Fig. 9a-f). The spatial distributions of rainwater mixing ratio are shown in Figure (9g-l). For Case-1 the NTU-3M simulates much higher rainwater mixing ratio than Morisson-2M. For Case-2 and Case-3 also in some areas over the domain NTU-3M simulates more rainwater mixing ratio than Morisson-2M. This is consistent with the time-height profile of rainwater mixing ratio (Fig. 6). The hydrometeors are averaged over land only and ocean only points of the respective domains and cases to quantify the difference in simulating the cloud hydrometeors (Table 4). This further highlights that the NTU-3M simulates more ice and lesser graupel and snow. Therefore, the results of cloud hydrometeors (cloud ice, snow, graupel, and rain mixing ratio) highlight the importance of choice of microphysical parameterization schemes. Significant differences in the water vapor mixing ratio are not noticed (Fig. S3). The water vapor is the initial source, coming from the initial/boundary conditions (IC/BC) and cumulus parameterization (Hazra et al., 2022). In this study, the IC/BC is kept same for experiments with different microphysical parameterization schemes for each case.

Cumulus parameterization was turned off. Hence experiments have not shown insignificant differences in the water vapor. The two different microphysical parameterizations produce different cloud ice mixing ratios. Variations in cloud ice formation contribute to fluctuations in the mixing ratios of mixed-phase cloud hydrometeors (snow and graupel) and rainwater.

A contoured frequency by altitude diagram (CFAD) is presented from both the microphysical parameterization schemes for each case in Figure 10. It is already seen that rainwater mixing ratio is higher in NTU-3M scheme which is consistent with the presence of more frequent high reflectivity (>30 dbZ) zones. Previous studies have also confirmed that the zone of maxima of reflectivity coincides with the maxima of precipitable water (Halder and Mukhopadhyay, 2016) in thunderstorm events. Proper modulation of CFAD profile of reflectivity case to case basis is absent in Morrsion-2M (Fig. 10a, b, c) scheme which is prominent in NTU-3M scheme (Fig. 10d, e, f). For example, in NTU-3M scheme, higher reflectivity bands are more frequent in Case-1 (Fig. 10d) near the surface than that of Case-2 (Fig. 10e) and Case-3 (Fig. 10f). Lightning flash counts observation (Fig. 2) from LDN detects more lightning flashes in Case-1 than the other two cases. However, Morrison-2M simulates similar profile across all the cases. Time-height profile of reflectivity (Fig. 11) averaged over lightning prone zone for each case also clearly shows the difference in simulation of reflectivity between these two microphysical schemes. Higher reflectivity is seen in the lower tropospheric levels from NTU-3M for Case-1 than Morrison-2M, which is in contrast with other two cases. The peak of reflectivity simulated is also in line with the diurnal variation of observed lightning flash counts (Fig. 5d).

### 3.3. LPI and rainfall:

As the NTU-3M, does not provide quantitative lightning flashes we could not compare the simulated lightning flashes between Morrison-2M and NTU-3M. Hence, we compared the simulated LPI with the lightning observation to qualitatively assess the performance of this scheme. Previous studies (Lagasio et al., 2017, Yair et al., 2010) have found a strong correlation between observed lightning and LPI. The Spatial distribution of LPI from both the schemes for each case is shown in Figure 12. From observation (Fig. 2) we have seen that lightning flashes over the domain on the eastern coast are more in Case-2 than Case-1. NTU-3M can correctly capture this variation in LPI between Case-1 (Fig. 12b) and Case-2 (Fig. 12d). Contrastingly, Morrison-2M shows more LPI for Case-1 (Fig. 12a) than Case-2 (Fig. 12c) over majority of the areas. Lightning flashes are also more observed in Case-3(Fig. 1c) than Case-1 (Fig. 1a). NTU-3M also shows realistically more LPI in Case-3 (Fig. 12f) than that of Case-1 (Fig. 12b). LPI is averaged over lightning zones of large activity as seen from the observation (Fig. 2) for each case and tabulated as Table 2. The rainfall values over those zones are also listed (Table 3). It also clearly quantifies that Morisson-2M simulates high LPI in Case-1 which is in sharp contrast with that of NTU-3M. Quantitative comparison of LPI values for NTU-3M also confirms that LPI simulation is consistent with lightning observation for all the three cases. LPI is dependent on the distribution of cloud hydrometeors. Hence, this reveals that the choice of 'moment' is playing a crucial role. In microphysical parameterization the 'moment' controls the interconversion of cloud hydrometeors. Therefore, proper choice of microphysical parameterization scheme e.g. NTU-3M may lead to realistic modulation of hydrometeors which in turn result in better lightning forecasting.

To understand the fidelity of simulating convection, rainfall production by both the schemes is also evaluated with the observation (TRMM) for all the cases. Among all the cases, highest rainfall is observed in Case-1 followed by Case-2 and Case-3 (Fig. 13a,d,g and Table 2). For Case-1 and Case-2, more rainfall is observed over land than ocean which is in contrast with Case-2. Though both the schemes can capture the overall rainfall spatial distribution there is area specific bias of rainfall for both the schemes (Fig. 13b,e,h and Fig. 13c,f,i and Table 3). To understand it quantitatively we have computed the mean statistics of the rainfall from observation and two sensitivity experiments for each case (Table 3). Morrison-2M shows the highest rainfall in Case-1. However, it simulates more rain in Case-3 than in Case-2, which is opposite to the observation. This inconsistency is improved in NTU-3M. The results also show that Morrison-2M simulates more (less) rainfall over land (ocean) for all the cases. On the other hand, NTU-3M shows more (less) rainfall over land (ocean) for Case-2 and Case-3. We have also computed the mean bias of rainfall for both the schemes over the whole domain along with land and ocean separately. The magnitude of bias is less in NTU-3M for most of the instances (Table 2). Both experiments show high dry bias over the ocean for Case-2. NTU-3M (Morrison-2M) shows high wet bias over the ocean for Case-1(Case-3). For all three cases, the bias of rainfall over land is less in NTU-3M than in Morrison-2M (Table 2, Fig. 13). Hence, the results imply that choice of microphysical parameterization scheme can influence the convection, evolution of hydrometeors and therefore rainfall. The results also highlight the importance of NTU-3M in the WRF model for better simulation of lightning related rainfall.

### 3.4. Drop Size Distribution:

Surface rainfall comprises a diverse spectrum of raindrops, characterized by mean drop size and droplet number concentration, which effectively represent rainfall-related features like convective type and associated atmospheric conditions. Therefore, understanding these characteristics is crucial for enhancing remote sensing capabilities and modeling rainfall phenomena (Ryu et al., 2021). The raindrop size distribution (RDSD) is the distribution of the number of raindrops according to their diameter (D), which is important for microphysical processes to account for the formation of rain droplets. Previous research (Chapon et al., 2008; Liao et al., 2014; Nelson et al., 2016) utilizing remote sensing data has highlighted the significance of drop size distribution (DSD) in rainfall for estimating rain intensity and adjusting latent heating profiles, a critical aspect for parameterizing rain microphysics in numerical weather forecasting models (Lim & Hong, 2010; G. Zhang et al., 2006, 2008). Barnes and Houze (2015) found that the development and lifetime of convective systems depends on ice phase microphysical process. Proper simulations of microphysical processes are also essential for DSD (Li et al., 2010, Iguchi et al., 2012, Chen et al., 2021). Hence to understand the impact of double and triple moment microphysical parameterization in this connection we have computed the DSD of several hydrometeors e.g. ice, graupel, snow and rain for both the schemes and all the cases (Fig. 14). Distinct difference in DSD of hydrometeors notably on cloud ice and raindrop is noticed between Morrison-2M and NTU-3M for all the cases. The Morrison-2M critically underestimates the bigger size of ice particles which is present in NTU-3M (Fig. 14a, c, i). Morrison-2M shows higher number density of snow than NTU-3M (Fig. 14b, f, j) which is consistent with spatial distribution of snow (Fig. 7 g-l). The slightly higher number density of graupel is also noticed for Morrison-2M than in NTU-3M. The DSD of raindrop is almost similar for both the schemes; however, bigger raindrops (> 2.5 mm) are more simulated in NTU-3M. For the case-1, more rain droplets are observed for all rain bin sizes Morrison-2M scheme as compared to NTU-3M scheme (Figure 14), which might be the reason for the overestimation of rainfall in Morrison-2M scheme for the case-1 (Figure 13). On the other hand, more (less) number of smaller (bigger) rain droplets are observed in NTU-3M scheme as compared to Morrison-2M scheme for Case- 2 and 3 (Figure 14).

## 4. Summary:

This study focuses on three specific lightning events to conduct a sensitivity analysis, aiming to understand the differences in simulating hydrometeors between double and triple moment microphysics schemes. The spatial distribution of lightning flashes revealed widespread activity over the head Bay and neighboring land mass for the first two cases and intense lightning over the southern peninsula for the third case.

The dynamics and distribution of cloud hydrometeors were examined, showing differences in vertical velocity and cloud hydrometeor profiles between the two schemes for all cases. Morrison-2M consistently simulated higher vertical velocity than NTU-3M. NTU-3M simulated more cloud ice, while Morrison-2M exhibited higher snow and graupel mixing ratios. Differences in rainwater mixing ratio were observed, with NTU-3M simulating more in some areas for all cases. The inclusions of ice crystal shapes in the NTU-3M are responsible for many of the key differences between these two microphysics simulations. The study emphasizes the importance of choice of microphysical parameterization schemes in simulating cloud hydrometeors.

Contoured Frequency by Altitude Diagrams (CFAD) highlighted the differences in reflectivity profiles between the schemes, with NTU-3M showing modulation of reflectivity profiles in association with the characteristic of the convection, unlike Morrison-2M. Lightning Potential Index (LPI) spatial distribution from both schemes was compared with observed lightning patterns, revealing NTU-3M's consistency with observations, while Morrison-2M showed discrepancies. Mohan et al. (2021) and Vani et al. (2022) have shown the higher false alarm ratio (FAR) in WRF with Morrison-2M microphysics, which is related to the overestimation of lightning flash counts as compared to LDN observation. Here, we have demonstrated that NTU-3M microphysics able to reduce the overestimation of lightning significantly for the thunderstorm event (Case-1) associated with an ESCS, which will definitely help to reduce FAR in operational forecasting.

This study also evaluated the rainfall simulation, noting that NTU-3M demonstrated improvements over Morrison-2M in capturing observed rainfall patterns, especially in terms of bias over land. Comparing Morrison-2M and NTU-3M schemes reveals differences in hydrometeor size distribution, with Morrison-2M underestimating larger ice particles. While raindrop distribution is similar, NTU-3M simulates larger raindrops. The results underscored the significance of moment of the microphysical parameterization scheme, specifically NTU-3M, for more accurate simulation of lightning and lightning related rainfall in the WRF model.

**Acknowledgement**

We acknowledge all the support received from MoES, the Government of India, and Director IITM to carry out this work. We are thankful to the team of LDN, IITM for the datasets used here. The Grid Analysis and Display System (GrADS), NCAR Command Language (NCL), Ferret-NOAA, Climate Data Operators (CDO), Python (https://www.python.org/) are also acknowledged for producing the results.   AH and MB are thankful to Director, and Project Directors of IITM, for providing encouragement to carry out the research work. AH also dully acknowledges National Science and Technology Council (NSTC), Taiwan for funding support as visiting researcher in National Taiwan University, Taiwan. UD is also thankful to National Science and Technology Council (NSTC), Taiwan for funding support.

**Data Availability Statement**

All data used in this study are available on request.

**Declaration of Competing Interest**

All authors declare that they have no known competing financial interests or personal relationships that could have appeared to influence the work reported in this paper.

**CRediT authorship contribution statement**

**Chandrima Mallick:** Methodology, WRF model simulations and analysis, Preparation of all data, Analysis, Writing -review & editing, Writing - original draft.

**Ushnanshu Dutta:** Analysis, Writing -review & editing, Writing - original draft.

**Moumita Bhowmik:** Analysis, Writing -review & editing, Writing - original draft.

**Greeshma M. Mohan:** WRF model simulations and analysis, Writing -review & editing, Writing - original draft. **Anupam Hazra:** Conceptualization, Guiding, Methodology, WRF model data post-processing and Analysis, Writing - review &

editing, Writing - original draft. **S. D. Pawar:** Observation data, LDN, Writing – review & editing, Writing - original draft. **Jen-Ping Chen:** Guiding, Methodology, WRF model, Writing - review & editing, Writing - original draft.